\documentclass[aip,jcp,preprint]{revtex4-1}

\usepackage{float}
\usepackage{graphicx}
\usepackage{color, soul}
\soulregister\cite7
\soulregister\ref7

\begin{document}

%\preprint{}

\title{Thermal conductivity calculation in anisotropic crystals by molecular dynamics : application to $\alpha\textrm{-Fe}_2\textrm{O}_3$} %Title of paper



\author{Jonathan Severin}
\email{jseverin@univ-montp2.fr}
%\thanks{}
\affiliation{Institut Charles Gerhardt Montpellier, UMR 5253 CNRS-UM-ENSCM, Place E. Bataillon CC1506 34095 Montpellier, France}
\affiliation{Total Research Center, Chemin du Canal BP 22 69360 Solaize France}

\author{Philippe Jund}
%\thanks{}
\affiliation{Institut Charles Gerhardt Montpellier, UMR 5253 CNRS-UM-ENSCM, Place E. Bataillon CC1506 34095 Montpellier, France}


\date{\today}

\begin{abstract}
In this work, we aim to study the thermal properties of materials using classical molecular dynamics simulations and specialized numerical methods. We focus primarily on the thermal conductivity $\kappa$ using non-equilibrium molecular dynamics to study the response of a crystalline solid, namely hematite ($\alpha\textrm{-Fe}_2\textrm{O}_3$), to an imposed heat flux as is the case in real life applications. We present a methodology for the calculation of $\kappa$ as well as an adapted potential for hematite. Taking into account the size of the simulation box, we show that not only the longitudinal size (in the direction of the heat flux) but also the transverse size plays a role in the determination of $\kappa$ and should be converged properly in order to have reliable results. Moreover we propose a comparison of thermal conductivity calculations in two different crystallographic directions to highlight the spatial anisotropy and we investigate the non-linear temperature behavior typically observed in NEMD methods.
\end{abstract}

%\pacs{}% insert suggested PACS numbers in braces on next line

%Use showkeys class option if keyword
\maketitle %\maketitle must follow title, authors, abstract and \pacs

\section{Introduction}
\label{sec:intro}

From {\em ab initio} calculations of material properties to finite-element models of macroscopic systems, numerical studies of the thermal transport are motivated by the wide range of technological applications. Examples of such applications are found in nanoelectronics \cite{sinha2005review}, aerospace \cite{lee2012large}, automotive \cite{yang2005potential} and building sector \cite{jelle2011traditional}. Therefore, being able to model the heat transfer and predict thermal properties presents a definite advantage for designing or improving materials and industrial processes.
\\
NEMD methods have been applied to thermal conductivity calculations since the beginning of the 1980s \cite{evans1982homogeneous} and a number of different algorithms have been developed \cite{schelling2002comparison}. Two main approaches can be identified: imposing the temperature gradient or imposing the heat flux. The latter approach has the advantage of a faster convergence \cite{heino2003thermal} and two of its algorithms are now commonly applied \cite{schelling2002comparison}: the particle velocity exchange presented in ~\cite{muller1997simple} and the heat source and heat sink method described in ~\cite{jund1999molecular}, ~\cite{ikeshoji1994non} and ~\cite{aubry2004robust}. Here we make use of the source and sink method which was first developed for amorphous materials and simple Lennard-Jones (LJ) liquids and has been generalized to simple crystals, e.g. FCC Cu, Ag, Au, etc. in ~\cite{bracht2014thermal, ge2013vibrational, zhang2013thermal, zhou2007phonon} and, in some cases, to more complex structures such as tetragonal ZrO$_2$\cite{schelling2001mechanism} or zeolitic imidazolate frameworks in ~\cite{zhang2013zeolitic}.
\\
The work reported here is part of a wider effort which aims to predict the material-lubricant interactions involved in the operation of a car engine by determining numerically the thermal conductivity of the different parts. As such, we focused in a first step on developing a methodology that we then applied to iron oxide (hematite) as a model material for the surface of the solid parts of the engine and we considered a temperature range between 300 and 500 K. In addition to its industrial and technological importance \cite{dzade2014density}, in recent years hematite has been subject to a renewed interest due to discoveries concerning the geological structure and mineral properties of Mars \cite{christensen2000detection, morris2006mossbauer, fraeman2013hematite}. Bulk hematite ($\alpha\textrm{-Fe}_2\textrm{O}_3$), of space group $R\overline{3}c (167)$ , has a crystal structure that can be indexed as hexagonal with a unit cell consisting of 6 formula units in which the oxygen ions lie approximately in a hexagonal close-packed framework while the iron ions are positioned symetrically in two-thirds of the octahedral interstices \cite{morrish1994canted, nolze2014exploring}. We chose to model this material using an empirical interatomic potential that we adapted and tested.
\\
The main objective of the study was to put together a set of methods and tools for the calculation of the thermal conductivity of a crystal based on a previous work on amorphous materials by non equilibrium molecular dynamics (NEMD) \cite{jund1999molecular}. We considered the influence of sample size, temperature and crystallographic orientation and we present a step by step methodology.
\\
The paper is divided into three sections. In the first section we describe the NEMD algorithm and methodology, the empirical potential used to model the material and the numerical methods applied to circumvent sample size effects. In the second part we present and discuss our results on the thermal conductivity of hematite single crystals, its temperature dependence, the spatial anisotropy and the nonlinear behavior observed in specific regions of the simulation box. Finally, we draw the major conclusions in the last section.

%------------------------------------------------

\section{Numerical methods}

	\subsection{Molecular dynamics and NEMD}

Classical molecular dynamics (MD) is a very popular method providing atomic-scale information on material properties and processes. Based on the integration of the Newton equations of motion at the atomic level, it allows to use relatively large simulation boxes compared to first-principles calculations. In non equilibrium molecular dynamics (NEMD) one or several internal variables are constrained to keep the system out of the equilibrium state. In our case we use the standard velocity-Verlet algorithm for time integration with a time step of 0.6~fs and we apply the method described in ~\cite{jund1999molecular} for the determination of the thermal conductivity $\kappa$ . The first step is choosing the direction of the heat flux. Here we will start by considering the z direction, parallel to the [001] crystallographic direction in the hexagonal lattice of hematite. Then we define two plates positioned at 1/4 and 3/4 of the simulation box length and orthogonal to the z axis. Periodic boundary conditions (PBC) are applied in every direction. At each time step, one of these plates receives a fixed amount of energy $\Delta \epsilon$ while the same amount is subtracted from the other. This is done by constantly rescaling the velocities of the atoms within the plates. This effectively results in imposing a heat flux across the box and parallel to the z direction. Care is taken to prevent a drift of the center of mass caused by the velocity rescaling. For a particle $i$ in the hot plate $P_+$, the rescaling can be expressed as 

\begin{equation}
\label{eq:rescal}
\mathbf{\overline{v}_i} = \mathbf{v_G} + \alpha (\mathbf{v_i} - \mathbf{v_G})
\end{equation}

where $\mathbf{\overline{v}_i}$ is the updated velocity, $\mathbf{v_G}$ the velocity of the center of mass of the ensemble of particles in $P_+$ and

\begin{equation}
\alpha = \sqrt{1 + \frac{\Delta \epsilon}{E_{P_+}}}.
\end{equation}

Of course, in the case of the cold plate $P_-$, $\Delta \epsilon$ should be subtracted. The non-translational kinetic energy $E_{P_+}$ is given by

\begin{equation}
E_{P_+} = \frac{1}{2} \sum_i m_i \mathbf{v_i}^2 - \frac{1}{2} \sum_i m_i \mathbf{v_G}^2.
\end{equation}

\begin{figure}[h]
\centering
\includegraphics[width=8cm]{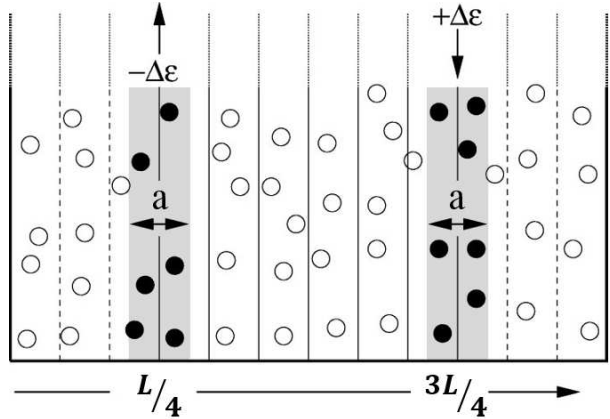}
\caption{Diagram of the NEMD method. The temperature slices are represented along with the heat source and sink (gray slabs). Adapted with permission from~\cite{jund1999molecular}. Copyrighted by the American Physical Society.}
\label{method}
\end{figure}

\begin{figure}[h]
\centering
\includegraphics[width=10cm]{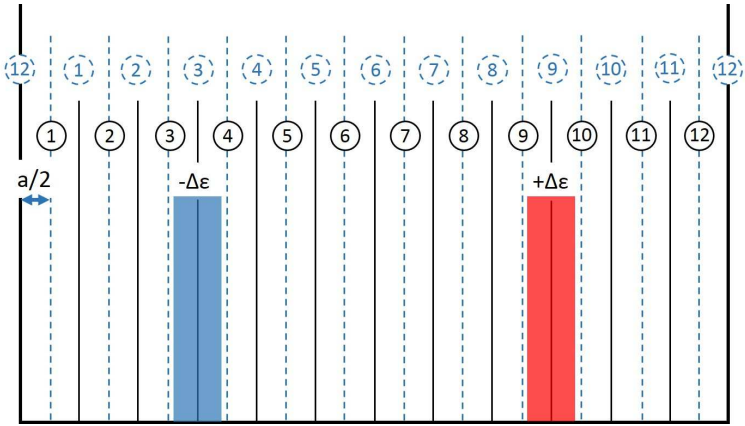}
\caption{Diagram of the NEMD method with the addition of a second set of temperature slices (dotted lines) translated by a/2 from the first.}
\label{overlap}
\end{figure}

To keep track of the temperature profile along the z axis, the box is divided into a set of slices in which the local temperature is computed, as summarized in figure \ref{method}. The width $a$ of those slices, and therefore the number of atoms in each of them, affects the temperature calculation. A larger number of atoms provides better statistics and so allows to reduce the time over which the values need to be averaged. But to obtain a detailed temperature profile the number of slices has to be large enough and therefore their width is limited. In order to eliminate this constraint we introduce several sets of overlapping bins, as described in figure \ref{overlap}. This allows to increase the temperature profile resolution without decreasing the number of atoms per slice. We find that 4 sets of 12 slices is a good compromise between precision and computational effort.

The initial state of the simulated system is obtained by replicating a smaller cell previously optimized by energy minimization. After a first run of relaxation in a (NPT) ensemble at the desired temperature, the heat flux is switched on. We have checked that small variations of the amount of energy added to or subtracted from the plates don't significantly affect the value of $\kappa$. Nevertheless, large values of $\Delta \epsilon$ will cause a large temperature difference between the source and the sink. A value of 1.5~\% of $k_B T$ seems reasonable for a simulation box containing 60000 atoms. Once the heat flux is switched on, it is necessary to let the dynamics run long enough to produce a steady state. Typically, this stationarization is of the order of 1 ns as reported in ~\cite{schelling2002comparison,maiti1997dynamical} for Stillinger-Weber silicon. We observe an average time of 0.8~ns before stable temperatures are observed in the source and sink plates. Then we start collecting and averaging the temperature values in every bin. Depending on the size of the system, we find that the averaging time necessary to remove the numerical noise and obtain a smooth temperature profile, on which a linear function can be fitted, varies from 2 to 6~ns. Except for the first relaxation run, all the calculations are performed in a (NVE) microcanonical ensemble. Finally, the temperature gradient is calculated from the time-averaged temperature profile and the Fourier's law (\ref{eq:Fourier}) is used to obtain $\kappa$. Recently, an adapted version of the heat exchange algorithm was proposed by Wirnsberger et al. \cite{wirnsberger2015enhanced} to correct a large drift of the total energy of the system observed in long-term simulations. However, in the present work we observe a moderate difference between the initial and final values of the total energy (of the order of $10^{-4}$~\%). This is due to the length of our simulations which is less than 10~ns and to the relatively small value of the time step chosen to ensure energy conservation.

\begin{equation}
\label{eq:Fourier}
\vec{J} = - \kappa \vec{\nabla}T
\end{equation}

The application of such a method takes its toll on numerical performance. Indeed the permanent temperature rescaling in the heat source and heat sink and the local temperature computation in each slice are time consuming. And with the constraints of stationarization and averaging, the time cost is a primary issue. Some of the simulations performed as part of this work required more than 100000 CPU hours individually. All molecular dynamics simulations were conducted using a modified version of the LAMMPS package \cite{plimpton1995fast} (based on version may 2015).

\subsection{Interaction potential}
\label{sec:potential}

To perform molecular dynamics on an ionic crystal such as hematite one needs to apply an interatomic potential defined to model the structure of the material as well as its physical properties. As mentioned in section \ref{sec:intro}, we consider a hexagonal unit cell with 6 formula units (30 atoms) and the following lattice constants, respectively a, b and c: $5.039$, $5.039$, $13.77$ \AA \cite{sadykov1996effect}.
We use a modified version of the potential $U(r)$ published in 2006 by Pedone et al. \cite{pedone2006self} to model our material. Three terms contribute to the equation of $U$: a short-range Morse potential, a short-range $r^{-12}$ repulsion and a long range Coulomb interaction. A cutoff distance of 7~\AA~was applied to the Morse and short-range repulsion terms and the Coulomb interaction was implemented by way of an Ewald summation. 

\begin{eqnarray}
\label{eq:potential}
U(r) = D_{ij} ([1 - e^{-a_{ij} (r - r_0)}]^2 - 1) + \frac{C_{ij}}{r^{12}} + \frac{z_i z_j e^2}{r}
\end{eqnarray}

The parameters of the potential were initially fitted on the experimental lattice constants, atomic positions and elastic constants using free energy minimization \cite{catlow1997computer} and other empirical fitting methods implemented in the GULP package \cite{gale1997gulp, gale1996empirical, richard1994self}. These methods allow to fit the potential on a structure relaxed at a finite temperature taking into acount quantities such as mechanical and dielectric properties. Nevertheless the experimental lattice constants used for this fit are for some reason different from the values obtained in most of the studies \cite{sadykov1996effect, sorescu2004hydrothermal, sawada1996electron, barinov1995effects, morrish1994canted, maslen1994synchrotron}. A fine-tuning of the Morse equilibrium parameter ($r_0$) for the Fe-O and the O-O interactions allowed us to accurately fit the ``correct'' experimental values. Indeed differences between lattice constants and interatomic distances experimentally observed and obtained with the modified potential are well under 1\%, as presented in table \ref{table:cellTab}. The new values of $r_0$ are 2.41810~\AA~and 3.65455~\AA~respectively for the Fe-O and O-O interactions.

\begin{table*}
\caption{Crystal structure validation}
\begin{ruledtabular}
\begin{tabular}{lllll}
%\toprule
Cell parameters. & Experimental (\AA) & Calculated.\footnote{calculated values from \cite{pedone2006self}} (\AA) & Calculated\footnote{this work} (\AA) & Difference (\%)\\
\hline
a   & 5.039 & 4.95  & 5.066 & 0.54 \\
b   & 5.039 & 4.95  & 5.066 & 0.54 \\
c   & 13.77 & 13.42 & 13.74 & 0.23 \\
c/a & 2.73  & 2.71  & 2.71  & 0.77 \\
%\bottomrule
\end{tabular}
\end{ruledtabular}
\label{table:cellTab}
\end{table*}

In addition to the structural properties, the bulk modulus was investigated in two different ways in order to validate further the modified potential function. First the elastic constants were calculated by deforming the hexagonal box in the 6 degrees of freedom and observing the change in the stress tensor at zero temperature. The Voigt-Reuss approximation applied to the calculated elastic tensor gives a value of 217~GPa for the bulk modulus. Additionnally, we performed single-point energy calculations at different volumes around the equilibrium volume of the cell and fitted the curve E=f(V) with the Vinet equation of state \cite{vinet1987compressibility}. We obtained this way a bulk modulus of 219~GPa. These calculated results fall within the measured values of 203 \cite{liebermann1968elastic} and 230~GPa \cite{catti1995theoretical} available in the literature. The small difference between the two calculated values (less than 1\%) may be used as an indication of the precision of this kind of calculations.

\subsection{Finite size effects}
\label{finiteSize}

With a band gap of 2.1 eV \cite{morrish1994canted}, the heat transport in hematite is primarily described by the lattice thermal conductivity which, in turn, is determined by the phonon transport. When performing NEMD on a finite simulation box, the phonon mean free path (MFP), which is the average distance a phonon travels before being scattered, is of particular importance. Phonons with different MFPs contribute differently to the thermal transport and in some materials the MFP of the contributing phonons can take values larger than several hundred nanometers \cite{sellan2010size, hermet2016lattice}. The NEMD method, as described in the previous sections, requires that two plates be defined at one quarter and three quarters of the box length to serve as heat source and sink. It would therefore be necessary to have a simulation box twice as long as the length of the largest contributing MFP. These scales are very difficult to handle in molecular dynamics simulations since the simulations would require enormous computing resources and very long execution times. 
We studied the effect of finite box dimensions in the direction perpendicular to the heat flux and we applied the method proposed by Schelling et al. in ~\cite{schelling2002comparison} to the finite size effects parallel to the heat flux. For the latter, several simulations with increasing box sizes $L$ are performed and their results are combined with the use of the so-called linear extrapolation procedure. As the size of the simulation box grows, more phonon mean free paths are taken into account which increases the calculated value of the thermal conductivity. As described in ~\cite{sellan2010size}, within a first order approximation a linear dependence can be expected between $1/\kappa$ and $1/L$. As a consequence, $\kappa_{\infty}$ can be evaluated by plotting $1/\kappa$ against $1/L$ and extrapolating the curve to $1/L_{\infty} = 0$ with a linear function. To obtain a reasonable precision of the curve to be fitted, a large number of individual simulations has to be performed. Nevertheless, the computational effort needed is considerably lower than for a real-size simulation with $L$ larger than the largest phonon mean free path. This method has also the advantage of providing a thermal conductivity value that relies on a significant number of different simulations with different initial states, thus reducing the statistical errors that could be attributed to an individual simulation.\\
Sellan et al. \cite{sellan2010size} mentioned a number of limitations of this method, due primarily to the first order approximation. For example, when considering a large collection of different box sizes, they reported good results for the application of the extrapolation method to Lennard-Jones argon, while an underestimation was observed in the case of Stillinger-Weber silicon. In the latter case, the calculated values and the extrapolated curve would diverge for the largest sizes. To address this issue, Hu et al. \cite{hu2011one} suggested a relation between the divergence and the aspect ratio of the simulation box (length/width). Studying Lennard-Jones solid argon, Lennard-Jones WSe$_2$ and graphite, their first conclusion is that a divergence can be observed at very high aspect ratios (200 to 300) for an elastically isotropic material such as LJ argon. Nevertheless these authors add that such high limits don't have practical consequences since a converged value of $\kappa$ can be computed well below those limits. Their second conclusion is that a clear divergent behavior is observed for low aspect ratios ($\sim$~30) for elastically anisotropic materials such as LJ WSe$_2$ and even earlier for graphite. 
With the smallest lateral size considered for the initial tests, we reached a maximum aspect ratio of 133. With the lateral size used in most of our simulations the largest aspect ratios range from 83 to 116. Since no divergence can be seen in the results presented hereafter, we can conclude that our material is probably elastically isotropic.
Moreover, Hu et al. proposed a criterion to predict if the thermal conductivity of a material can be modeled by NEMD methods or not. This criterion is based on the ratio of the elastic constants along two crystallographic directions ($\frac{C_{xx}}{C_{zz}}$) which should be reasonably low (\textless~5). We computed the 6x6 matrix of the elastic constants as described in section \ref{sec:potential} and obtained the following ratio:
\begin{equation}
\frac{C_{xx}}{C_{zz}} = \frac{362}{326} \sim 1.1
\end{equation}
which shows that our material is perfectly suited for NEMD methods and is not prone to the divergence issues described in \cite{hu2011one}.

%------------------------------------------------

\section{Results and discussion}

\subsection{Size-dependent simulations}
\label{sec:simulations}

\begin{figure}[h]
\centering
\includegraphics[width=10cm]{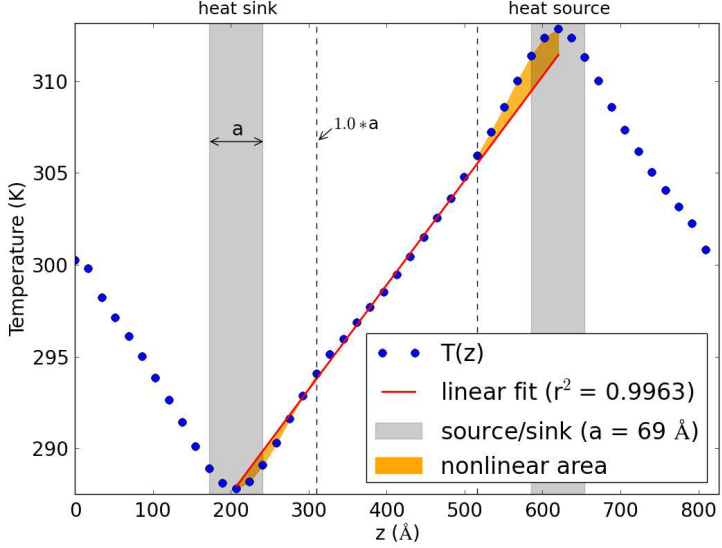}
\caption{Typical simulation result with the time-averaged temperature profile (blue dots) as a function of the position along the the z axis, parallel to the heat flux. The gray slabs are the areas corresponding to the heat sink and source. The dotted vertical lines are the limits of the linear domain where the slope of the curve is calculated with a fitting procedure.}
\label{typical}
\end{figure}

As explained in the previous section, the calculation of $\kappa$ requires two steps. First a number of size-dependent simulations, then a size-independent extrapolation. Figure \ref{typical} shows the typical result for an individual simulation for a given box size $L$. The time-averaged temperature profile is presented along with the position and width of the slabs where heat is added or subtracted (heat source and heat sink). In the central part of the simulation box, the temperature exhibits a linear profile which can be fitted to calculate the thermal gradient. Non-linear areas are noticeable close to the heat source and sink.

\begin{figure}[h]
\centering
\includegraphics[width=10cm]{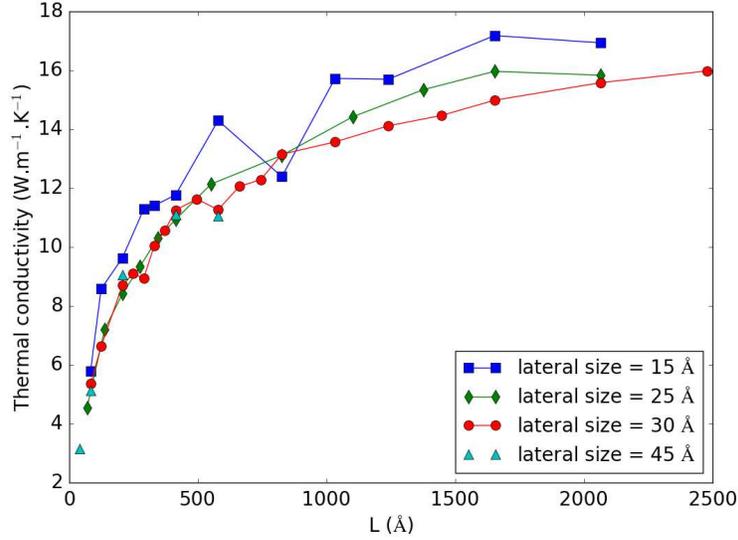}
\caption{Thermal conductivity at 300 K for different lateral box sizes as a function of the box size in the direction of the heat flux.}
\label{300size}
\end{figure}

In figure \ref{300size} the thermal conductivity at 300 K is shown as a function of the box size in the directions both orthogonal and parallel to the heat flux. One can observe a quick convergence of $\kappa$ as a function of the lateral size, and we fixed this size to 30~\AA~for the rest of the study. As for the evolution of $\kappa$ as a function of $L$, the size parallel to the heat flux (horizontal axis in figure \ref{300size}), it requires the application of the extrapolation method discussed in section \ref{finiteSize}. 

\begin{figure}[h]
\centering
\includegraphics[width=10cm]{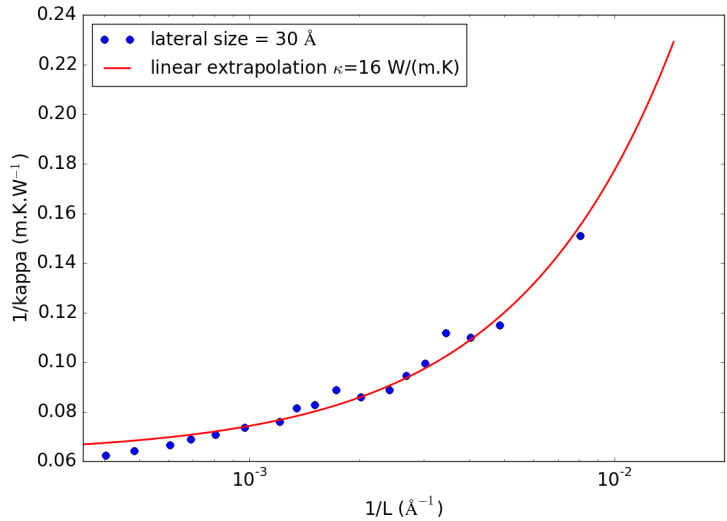}
\caption{The inverse of the thermal conductivity as a function of the inverse of the box size with a decimal logscale x-axis (blue dots). The line is the result of the extrapolation procedure.}
\label{extrapolated}
\end{figure}

A plot of $1/\kappa$ against $1/L$ is presented in figure \ref{extrapolated} showing the calculated values and the extrapolated curve with a logscale x axis. The corresponding value of $\kappa$ is calculated to be 16~W.m$^{-1}$.K$^{-1}$ for a pure single crystal of hematite at 300 K in the [001] crystallographic direction. Several thermal conductivity measurements of polycrystalline hematite at room temperature are reported in the litterature. The values presented in ~\cite{diment1988thermal, clark1966handbook, horai1971thermal} range between 11 and 13~W.m$^{-1}$.K$^{-1}$ while Akiyama et al. report a much larger value of 17~W.m$^{-1}$.K$^{-1}$ in ~\cite{akiyama1992measurement}. However, experimental values of $\kappa$ for single-crystal hematite samples are less common. In ~\cite{clauser1995thermal}, references are made to a 1974 work \cite{dreyer1974properties} where a value of 12.1~W.m$^{-1}$.K$^{-1}$ was reported for a single crystal in the [001] direction and 14.7 W.m$^{-1}$.K$^{-1}$ (22 \% more) for a second direction orthogonal to [001]. 

\subsection{Temperature dependence}

\begin{figure}[H]
\centering
\includegraphics[width=10cm]{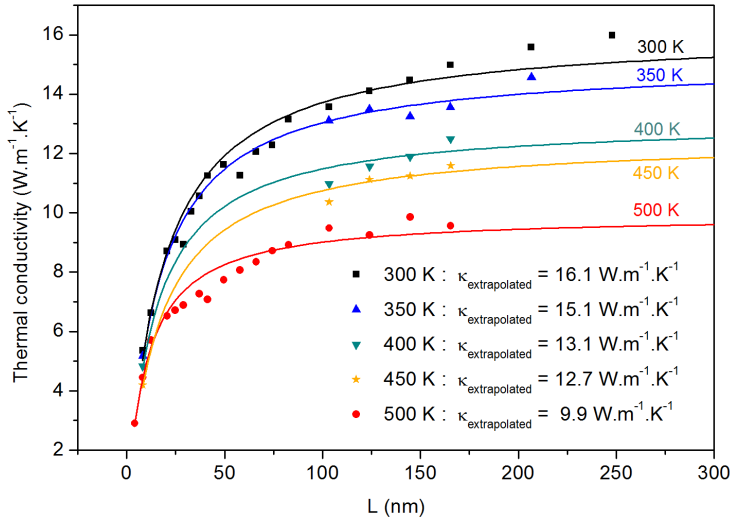}
\caption{Thermal conductivity as a function of box size for different temperatures along with the corresponding extrapolations (lines).}
\label{temperatures}
\end{figure}

In order to determine the temperature dependence of $\kappa$ in the range 300 to 500 K, it is in principle necessary to apply the previously described procedure for several temperatures. But taking into account the computational cost of such a brute force method, we propose an optimized approach where a full study is conducted at 300 K and 500 K but only a partial analysis is done in between.
Applying the full extrapolation procedure at the limits of the range provides a validation of its applicability. Assuming that the thermal conductivity follows the same behavior at the intermediate temperatures, we perform only a limited number of simulations at the lowest and highest box sizes for those temperatures providing thus the start and end points for the extrapolation. The results are coherent as can be observed in figure \ref{temperatures} and validate the approach. We estimate that proceeding in this way required 30 to 40 \% less computational time than a full study.

\begin{figure}[H]
\centering
\includegraphics[width=10cm]{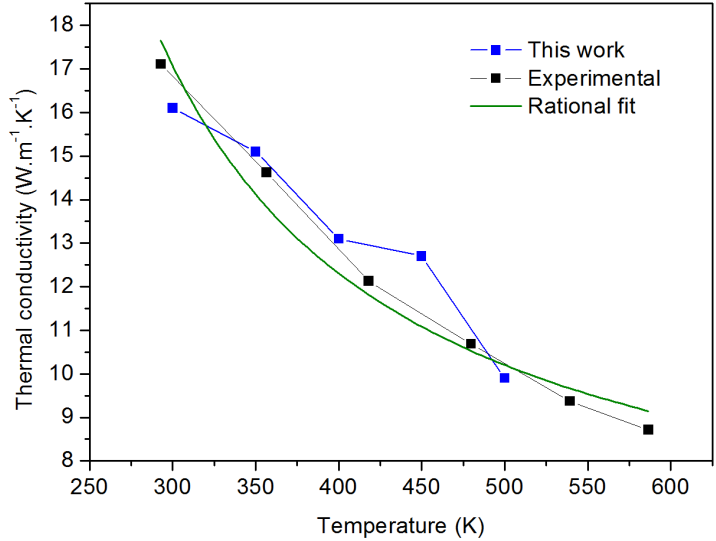}
\caption{Thermal conductivity as a function of temperature. The black squares are experimental values for polycristalline hematite from \cite{akiyama1992measurement}. These values were fitted with a function of the form $a/T+b$ to highlight the $1/T$ dependence (green line).}
\label{kappaTemp}
\end{figure}

From the infinite box size extrapolations, the temperature dependence of the thermal conductivity can be assessed. The results are presented in figure \ref{kappaTemp} along with a comparison with measurements made on polycristalline hematite in ~\cite{akiyama1992measurement}. As expected, $\kappa$ decreases with the temperature in the investigated range. In this temperature range it is reasonable to think that Umklapp processes are dominant and thus $\kappa$ should decrease like 1/T: this behavior is not obvious from the calculated values of figure \ref{kappaTemp}. However, we observed that the individual, size-dependent, conductivity values such as the ones shown in figure \ref{300size} may vary by up to 7.5\% when differences are introduced in the initial state of the simulations (e.g. random initial velocity distributions). And even though the extrapolated, size-independent, values of $\kappa$ rely on several different simulations, the lack of precision makes it difficult to assess a precise mathematical function from the curve of figure \ref{kappaTemp}.

\subsection{Spatial Anisotropy}

\begin{figure}[h]
\centering
\includegraphics[width=10cm]{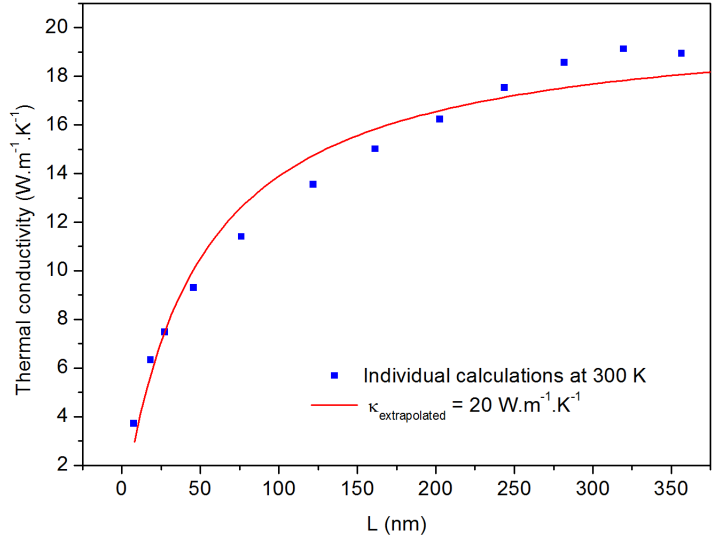}
\caption{Thermal conductivity in the [100] direction as a function of box size (blue squares). The line is the result of the extrapolation procedure.}
\label{extrapolated100}
\end{figure}

The results presented in the previous sections were obtained for a heat flux in the [001] crystallographic direction. To investigate the spatial anisotropy of $\kappa$, we applied the same methods to the [100] direction at 300 K. Figure \ref{extrapolated100} shows the evolution of $\kappa$ as a function of system size together with the extrapolated curve. A spatial anisotropy is observed for hematite since we obtained a value of 20 W.m$^{-1}$.K$^{-1}$ for $\kappa$ with the heat flux orientated in the [100] crystallographic direction. This is 25 \% more than in the [001] direction which is consistent with the experimental values mentioned earlier in section \ref{sec:simulations}.

\subsection{Non-linearity}
\begin{figure}[h]
\centering
\includegraphics[width=16cm]{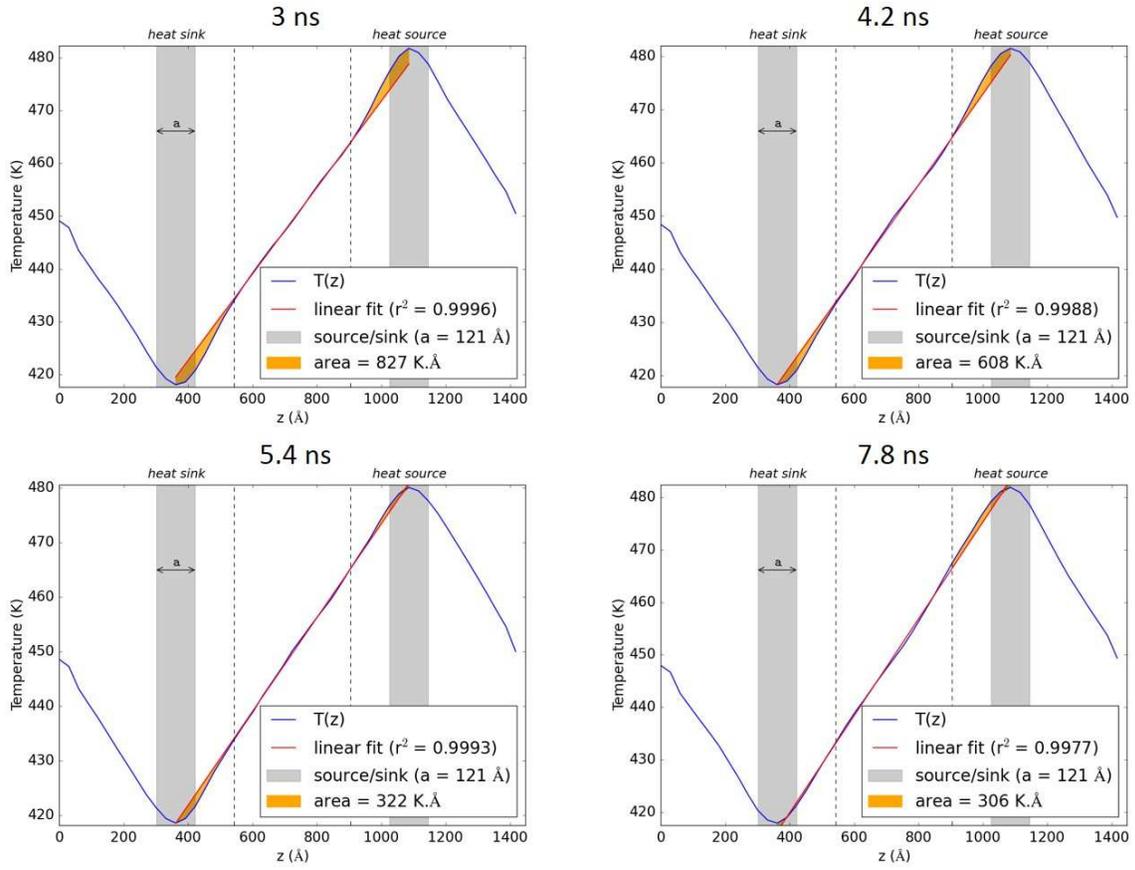}
\caption{Evolution of the time-averaged temperature profile for a 140~nm long simulation box made of 3780 hematite unit cells (113000 atoms) during an 7.8~ns long simulation. The yellow-colored areas between the temperature curve and the linear regression are computed to quantify the non-linearity.}
\label{prof_vs_temp}
\end{figure}

As can be observed in figure \ref{typical}, the temperature profile exhibits a nonlinear behavior near the source and the sink. This behavior has been partially explained by phonon scattering in previous studies \cite{schelling2002comparison, sellan2010size}. In order to calculate the temperature gradient and apply Fourier's law, it is thus necessary to consider the profile far enough from those non-linear areas. Our experience led us to choose an ``exclusion'' distance equal to the width $a$ of the temperature bins (heat source and sink included). Indeed, we found that this specific choice was appropriate in most of the simulations performed in this work, independently of the actual value of $a$. Moreover, investigating the evolution of the non-linearity as a function of time, we find that in the case of long enough simulations this non-linearity decreases significantly even after the system is considered to have reached the steady state. Figure \ref{prof_vs_temp} shows the evolution of the time-averaged temperature profile in a 140~nm long simulation box made of a 6x6x105 supercell (113000 atoms) during simulations lasting up to 7.8~ns. 

\begin{figure}[H]
\centering
\includegraphics[width=10cm]{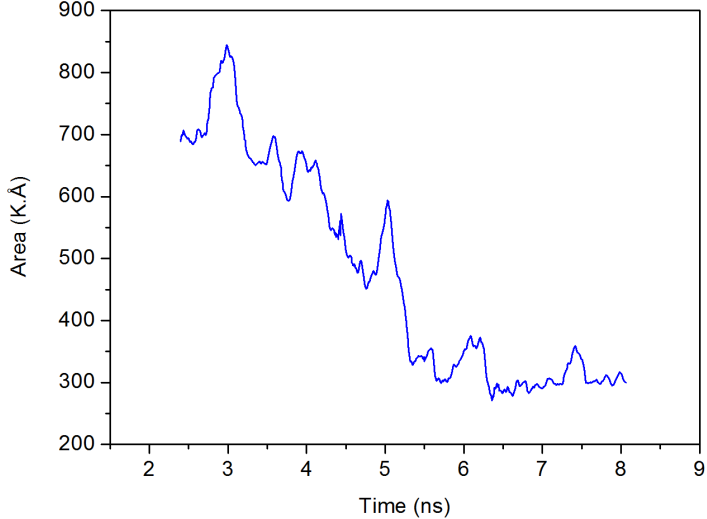}
\caption{Size of the non-linearity areas as a function of time from the end of the stationarization up to 8~ns.}
\label{nonLinear}
\end{figure}

The area between the temperature curve and the linear regression (colored area in figure \ref{prof_vs_temp}) was calculated to quantify the non-linearity. Figure \ref{nonLinear} shows the evolution of this quantity, starting after the stationarization. The value peaks around 3~ns and decreases afterwards, until a minimum value is reached. We have observed this evolution with time for different system sizes. Thus it appears that the non-linearity of the temperature gradient close to the source and the sink can partly be characterized as a slow transient phenomenon.

\subsection{Conclusion}
A detailed methodology has been presented for the determination of the thermal conductivity of crystals and applied to pure single-crystalline hematite. Calculated values for different temperatures are in reasonable agreement with available experimental data on polycrystals with values ranging from 16 to 10~W.m$^{-1}$.K$^{-1}$ between 300 and 500 K. Moreover, an investigation of the spatial anisotropy has been undertaken and shows, at 300~K, a 25\% increase of the thermal conductivity in the [100] direction with respect to the [001] direction in agreement with measurements on single crystals. Finally, some specific elements of the calculation procedure, such as the width of the temperature bins or the nature of the nonlinear behavior, have been analyzed highlighting new aspects in the application of the NEMD scheme for the determination of the thermal conductivity of crystalline solids.

\begin{acknowledgments}
This work was granted access to the HPC resources of CINES under the allocation 2016-c2016097598 made by GENCI. We gratefully acknowledge Total S.A. and Total M.S. for their financial support and we thank Sophie Loehl\'e and Vincent Lacour for fruitful discussions.
\end{acknowledgments}

% If in two-column mode, this environment will change to single-column format so that long equations can be displayed. 
%\begin{widetext}
%$$\mbox{put long equation here}$$
%\end{widetext}

\bibliography{biblio}

\end{document}